\pdfoutput=1
\documentclass[a4paper,11pt]{article}
\usepackage{tikz-cd}
\usepackage[dvipsnames]{xcolor}
\usepackage{amsmath,amssymb,amsthm,amsfonts,mathrsfs}
\usepackage{bm}
\usepackage{enumitem}
\usepackage{booktabs, comment, tikz}
\usepackage{tabularx}
\usepackage{jheppub}
\usepackage[T1]{fontenc}
\usepackage[utf8]{inputenc}
\usepackage{hyperref}
\usetikzlibrary{calc}
\hypersetup{
	linktoc=all,
	colorlinks=true,
	linkcolor=blue,
	citecolor=magenta,
	urlcolor=red
}

\newcommand{\dd}{\mathrm d}
\newcommand{\cE}{\mathcal E}
\newcommand{\cU}{\mathcal U}
\newcommand{\cW}{\mathcal W}

\newcommand{\cB}{\mathcal B}

\newcommand{\cG}{\mathcal G}

\usepackage[most]{tcolorbox}

\tcbset{highlight math style={left=02mm,right=02mm,top=02mm,bottom=02mm}} 
\usepackage{empheq}

\newcommand\inbox[1]{\tcbset{fonttitle={\mathcal I}ptsize} \tcboxmath[colback=white,colframe=black!70]{#1}}

\newcommand{\cM}{\mathcal{M}}
\newcommand{\cV}{\mathcal{V}}
\newcommand{\bM}{\partial_{_{\text{P}}}\mathcal{M}}
\newcommand{\pB}{{\cal B}_{_{\text{P}}}}

\makeatletter \@addtoreset{equation}{section}

\newcommand\bc{\begin{center}}
	\newcommand\ec{\end{center}}

\begin{document}
	
	\vspace*{-5cm}
	\title{\bc
		\huge{Null String Holography}\\
		\centerline{\LARGE{Null Strings Probe Projective Boundary of Their  Target Spaces}}
        \ec}
	
	\author[a]{M.M. Sheikh-Jabbari }
	\author[b]{, H. Yavartanoo}
	
	\affiliation[a]{School of Physics, Institute for Research in Fundamental Sciences (IPM),
		P.O. Box 19395-5531, Tehran, Iran}
	\affiliation[b]{Beijing Institute of Mathematical Sciences and Applications (BIMSA),
		Huairou District, Beijing 101408, P. R. China}
	
	\emailAdd{jabbari@theory.ipm.ac.ir}
	\emailAdd{yavar@bimsa.cn}

\abstract{We prove that the \textit{minimal null string theory} on a background manifold ${\cal M}$ is classically equivalent to the ILST theory on the  projective boundary of $\cM$, $\pB=\bM$. We conjecture that this equivalence also holds at quantum level, we dub this correspondence  \textit{null string holography}. In particular, we show the minimal null string theory on the Poincar\'e patch of a $D$ dimensional Anti de Sitter (AdS$_D$) background is described by the ILST theory on $(D-1)$-dimensional Minkowski background, paralleling the celebrated AdS/CFT. The minimal null string theory on the cosmological patch of dS$_D$ is governed by the ILST theory on $(D-1)$-dimensional flat Euclidean space. For null strings on a $D$-dimensional Minkowski space $\mathbb{M}_D$, depending on the coordinate patches used to span $\mathbb{M}_D$, the projective boundary includes a dS$_{D-1}$, asymptotic null infinities ${\cal I}^\pm$ and the hyperboloid $\mathbb{H}^{D-1}$ patches. The  ILST theory dual to the minimal null strings on each patch, then resides on the piece of the projective boundary spanned by the coordinates adopted in the patch. We comment on possible  applications of null string holography to black hole horizons. }

	\maketitle
	
\section{Introduction}

Null string theory, despite of half a century literature \cite{Schild:1976vq, Isberg:1993av, Bagchi:2026wcu}, was only recently given a formal definition which can be extended and generalized  \cite{Sheikh-Jabbari:2026kwr}. The definition in \cite{Sheikh-Jabbari:2026kwr}, parallels the same logic and argument of the Polyakov action \cite{Polchinski:1998rr} in defining tensile string theory: All worldsheets that are physically indistinguishable must be identified and the transformations mapping indistinguishable worldsheet geometries must be gauged. With the same token, a null string's worldsheet, which is a two-dimensional ($2d$) Carrollian geometry, are defined up to $2d$ diffeomorphisms, conventional Weyl and Carroll-Weyl scalings \cite{Sheikh-Jabbari:2026vqh}. Therefore, null strings should be described by an action in which all the three transformations are gauged. 

Nonetheless, historically, the null strings are defined by the celebrated ILST action \cite{Isberg:1992ia, Isberg:1993av} in which only $2d$ diffeomorphisms and conventional Weyl scaling are gauged. Thus, in our recent papers \cite{Sheikh-Jabbari:2026cnj, Sheikh-Jabbari:2026tpf, Sheikh-Jabbari:2026vqh} we raised the issue that ILST on its own is not describing a null string theory and should be supplemented with extra requirements that in effect reinforce gauging of the Carroll-Weyl scaling. We formulated the \textit{minimal null string theory} which has CW gauge symmetry by introducing the minimum needed extra worldsheet fields and requiring that added fields have a clear worldsheet geometric meaning. 

In this work, we ask the question: if not a null string theory, \textit{what does the ILST theory really describe?} To formulate this question, we consider ILST action, which is written on $D-1$ dimensional flat Minkowski background, and replace the Minkowski metric $\eta_{\mu\nu}$ with a generic nondegenerate metric $h_{\mu\nu}$ to obtain  a generic $D-1$ dimensional background geometry $\cB$ whose metric is $h_{\mu\nu}$. For nondegenerate (non-Carrollian) target spaces, we show that indeed this ILST action does describe a minimal null string theory on a $D$ dimensional background $\cM$ such that $\cB$ is its projective boundary. We dub this equivalence \textit{null string holography}. We exemplify the null string holography for some interesting cases, including $\cM$ being a $D$ dimensional flat Minkowski $\mathbb{M}_D$ and anti de Sitter (AdS$_D$) and de Sitter (dS$_D$) spacetimes. 

\paragraph{Note added.} While this work was in completion the paper 
\cite{Lindstrom:2026tua} appeared which discusses a similar problem for null strings on flat space. Their analyses and results are compatible with ours in section \ref{sec:Mink-bdry}.

\section{Minimal null string theory action}\label{sec:CW-reduction}

To set the conventions, let us start with the standard ILST action \cite{Isberg:1993av},
\begin{equation}
 S_{\rm ILST}:=S_{\rm ILST}[\eta; \cV]=
 \frac{\kappa}{2}\int_\Sigma \mathrm d^2\sigma\,
 \mathcal V^a\mathcal V^b \eta_{\mu\nu}
 \partial_aX^\mu \partial_bX^\nu.
 \label{eq:ILST-action}
\end{equation}
where, $\sigma^a=(\tau, \sigma)$ are the worldsheet coordinates, $\cV^a$ is a worldsheet vector density of weight $+1/2$  under diffeomorphisms, $X^\mu=X^\mu(\tau,\sigma)$ are $D$ dimensional embedding coordinates that span over the Minkowski background; $X^\mu$ are scalars under $2d$ diffeomorphisms and invariant under Weyl scaling. The ILST action is hence invariant under $2d$ diffeomorphisms and Weyl scaling.

The ILST action, is however, not invariant under Carroll-Weyl scaling. The minimal gauged ILST action on a $D$ dimensional flat Minkowski background,  which described by the \textit{minimal null string (MNS) action}, is \cite{Sheikh-Jabbari:2026vqh, Sheikh-Jabbari:2026kwr}
\begin{equation}
 S_{\text{MNS}}[\eta; \cV]
 =\frac{\kappa}{2}\int_\Sigma \mathrm d^2\sigma\,
 \mathcal V^a\mathcal V^b \eta_{\mu\nu}
 D_aX^\mu D_bX^\nu.
 \label{eq:CW-action-Mink}
\end{equation}
where
\begin{equation}
 D_aX^\mu
 :=\partial_aX^\mu+\mathcal W_a X^\mu.
 \label{eq:Dilaton-covariant-derivative}
\end{equation}
is the Carroll-Weyl (CW) covariant derivative, $\cW_a$ is the CW gauge-connection. This action is constructed such that it is invariant  under local CW scaling,
\begin{equation}\label{CW-scaling-infinitesimal-K=X}
    \delta_\chi \cW_a=-\partial_a\chi, \qquad \delta_\chi X^\mu =\chi X^\mu,\qquad \delta_\chi \cV^a=-\chi \cV^a\,,
\end{equation}
which yield the finite form of the CW scalings,
\begin{equation}\label{CW-scaling-finite-K=X}
    \cW_a\to \cW_a-\partial_a\chi, \qquad X^\mu \to e^{\chi} X^\mu,\qquad \cV^a\to e^{-\chi} \cV^a\,.
\end{equation}
More discussion and analysis about classical aspects of the above action may be found in \cite{Sheikh-Jabbari:2026vqh,Sheikh-Jabbari:2026tpf}.   

For comparison and later use, we note that the ILST action is the same as the action \eqref{eq:CW-action-Mink} without CW gauging, i.e. with setting $\cW_a=0$ and without \eqref{CW-scaling-infinitesimal-K=X} or \eqref{CW-scaling-finite-K=X} identifications. {On this nondegenerate target, the} ILST action does not realize the local CW transformations above.\footnote{We note, however, that when target space is Carrollian, the action can itself realize CW symmetry when the generator lies in its kernel and satisfies the required scaling condition \cite{Sheikh-Jabbari:2026nez}.}

The tensile string on a given background manifold $\cM$ with metric $g_{\mu\nu}(X)$ is described by the Polyakov action for flat Minkowski background, by simply replacing the Minkowski metric $\eta_{\mu\nu}$ with $g_{\mu\nu}$. This works because under $2d$ diffeomorphisms $X^\mu$ and hence $g_{\mu\nu}(X)$ behave as scalars and under conventional Weyl scaling $X^\mu$ do not change. The same simple replacement does not work for the minimal null strings described by \eqref{eq:CW-action-Mink}, because we now have CW scaling under which $X^\mu$ transforms as \eqref{CW-scaling-finite-K=X}. So, the first question to tackle is to construct minimal null string action on background $\cM$.

\subsection{Minimal null string on non-flat backgrounds}
Let $(\mathcal M,g)$ be a $D$-dimensional target space. The CW scaling, e.g. see \eqref{CW-scaling-finite-K=X}, suggests that the naive gauging, i.e. starting from ILST action, replacing $\eta_{\mu\nu}$ with $g_{\mu\nu}$ and replacing $\partial_a$ with a CW covariant derivative does not yield an action that is invariant under CW gauge symmetry, a minimal null string theory on background $g_{\mu\nu}$. As our analysis below makes clear, to be an admissible (classical) background for a minimal null string action,  $g_{\mu\nu}$ should have some scaling properties. As our analyses below make clear, {with} the {fixed} worldsheet {weight used below,} CW {invariance} {requires} {a} {vector} {satisfying} {\eqref{Dilatation-Eq};} {general} conformal Killing vectors (CKV) {are treated in section~\ref{sec:NMNS-K1-K2}}. In this section and to illustrate the basic idea we start with the assumption that $\cM$ (or a patch of it) admits a homothety vector field $\rho_\mu$ such that,
\begin{equation}\label{Dilatation-Eq}
    \nabla_\mu \rho_\nu+\nabla_\nu \rho_\mu =2 \ g_{\mu\nu},
\end{equation}
where $\nabla_\mu$ is the metric compatible covariant derivative w.r.t $g_{\mu\nu}$. The above equation implies that $\nabla^\alpha \rho_\alpha={D}$. We use metric $g_{\mu\nu}$ to raise and lower indices on tensors. 

{\paragraph{Important Notes:} 
\begin{itemize}[leftmargin=3pt]
    \item In general a given manifold $\cM$ or a patch of it need not admit a  homothetic vector field. There are a bigger class of manifolds (or patches thereof) which admit CKV. As we will discuss in \ref{sec:NMNS-K1-K2} null string theory can be defined for them too.
    \item If $\cM$ admits a homothety, $\rho^\mu\partial_\mu$ may not be globally defined and/or its causal character may change. That is, $\rho^2:= \rho_\mu \rho^\mu$ can be positive, negative or zero on different patches of the spacetime.
    \item In this work  we focus on patches of the spacetime in which $\rho^2$ does not change causal character.
    \item  Even in patches with a given causal character, the two $\rho^2=0$ and $\rho^2\neq 0$ cases should be discussed separately. 
    \item  Let us first focus on the $\rho^2\neq 0$ and in section \ref{sec:rho=0} we analyze {$\rho^2=0$ with $\rho\neq0$}. One can check that,
\begin{equation}\label{Lie-rho-h}
    {\cal L}_\rho h_{\mu\nu}=0,\qquad h_{\mu\nu}:= {\frac{1}{|\rho^2|}}\ g_{\mu\nu},
\end{equation}
because ${\cal L}_\rho (\rho^2)=2 \rho^2$, where ${\cal L}_\rho$ denotes Lie derivative along $\rho^\mu$. That is, $\rho^\mu$ is a Killing vector of the conformally scaled metric $h_{\mu\nu}$. 

{\item The choice of gauged vector $\rho$ is understood in the abbreviated action labels; it need not be uniquely fixed by the metric. We work locally away from its zeros, with a nonvanishing worldsheet density.} 

\end{itemize}

With the above requirement, one can readily verify that the gauged action
\begin{equation}
 S_{\text{MNS}}[g; \cV]
 =\frac{\kappa}{2}\int_\Sigma \mathrm d^2\sigma\,
 \cV^a\cV^b g_{\mu\nu}(X)
 D_aX^\mu D_bX^\nu\,.
 \label{eq:CW-g-cD}
\end{equation}
Throughout this work we assume the target space metric is nondegenerate, non-Carrollian; for Carrollian target spaces, see \cite{Sheikh-Jabbari:2026nez}. In the above,  
\begin{equation}\label{Da-cov-diff-X-cD}
    D_a X^\mu= \partial_a X^\mu + \cW_a \rho^\mu\,,
\end{equation}
is invariant under the local CW-scaling, generated by transformations,
\begin{equation}
 \delta_\chi X^\mu=\chi \rho^\mu(X),
 \qquad
 \delta_\chi\mathcal W_a=-\partial_a\chi,
 \qquad
 \delta_\chi\cV^a=-\chi\cV^a,\qquad \chi=\chi(\tau,\sigma)\,.
 \label{eq:intrinsic-CW-transformations}
\end{equation}
Invariance of the action may be verified noting that
\begin{equation}
 \delta_\chi (D_aX^\mu)=\chi (\partial_\nu\rho^\mu) D_a X^\nu\  ,
 \qquad
 \delta_\chi g_{\mu\nu}= (\partial_\alpha g_{\mu\nu}) \delta_\chi X^\alpha=\chi\rho^\alpha\partial_\alpha g_{\mu\nu},
 \label{eq:delta-chi-DX-g}
\end{equation}
and recalling \eqref{Dilatation-Eq}.  We note that the finite CW-scaling takes the form
\begin{equation}
\cV^a\to e^{-\chi}\cV^a,\qquad \cW_a\to \mathcal W_a-\partial_a\chi,\qquad X^\mu \to e^{\chi \rho^\nu\frac{\delta}{\delta X^\nu}}\ X^\mu.
\end{equation}
which for $\rho^\mu=X^\mu$ reduces to \eqref{CW-scaling-finite-K=X}.  As we see for a general metric $g_{\mu\nu}$ the $X^\mu$ transformation and its gauge-orbit identification is not linear in $X^\mu$.

\subsection{Connection to ILST theory}\label{sec:ILST-Bulk}

To see the connection of the minimal null string action $S_{\text{MNS}}[g; \cV]$ to ILST action which does not enjoy CW scaling as a gauge symmetry, we try to rewrite  \eqref{eq:CW-g-cD} in terms of CW gauge invariant quantities. This involves two steps, a change of coordinates on the target space and elimination of the CW connection $\cW_a$, that we will discuss below. 

\subsubsection{CW scaling invariant quantities}\label{sec:Inv-quantity-rho}

Given a Killing vector of the metric $h_{\mu\nu}$, it is natural to adopt a coordinate system where one of the coordinates is along the vector $\rho^\mu$:
\begin{equation}
 \rho^\mu\partial_\mu=\partial_s,\qquad \rho^2=e^{2s}\,.
\end{equation}
Here we assume $\rho^2 >0$ on the patch of spacetime we consider, $\rho^2<0$ case may be dealt with in a similar way by taking $\rho^2=-e^{2s}$. 
In this coordinate system and $h_{\mu\nu}$ is $s$-independent:
\begin{equation}\label{Lie-rho-h-Killing}
 g_{\mu\nu}(s,Y)=e^{2s}h_{\mu\nu},
 \qquad h_{\mu\nu}:=e^{-2s}g_{\mu\nu},
 \qquad \partial_s h_{\mu\nu}=0\,,
\end{equation}

The quantities appearing the MNS action can be written in the $(s, Y^i)$ basis. Let us start with,
\begin{equation}
 \qquad D_as=\partial_as+\mathcal W_a,
 \qquad D_aY^i=\partial_aY^i.
\end{equation}
Next, we introduce the invariant kernel vector $\cU^a$, 
\begin{equation}\label{cU-cV}
 \mathcal U^a:=e^s\mathcal V^a,
\end{equation}
Thus, we have
\begin{subequations}\label{rho2-delta-chi-inv}
\begin{align}
 \delta_\chi Y^i=0\,,\qquad \delta_\chi s=\chi,\quad & \quad {\delta_\chi \cW_a=-\partial_a \chi} \label{s-Wa-Y-chi} \\  \delta_\chi\mathcal U^a=0,
 \qquad \delta_\chi(D_as)=\partial_a\chi-\partial_a\chi=0, \qquad & 
  \delta_\chi h_{\mu\nu}=\delta_\chi s\ \partial_sh_{\mu\nu}=0\,.
\end{align}
\end{subequations}
Using $\mathcal V^a\mathcal V^b g_{\mu\nu}=\mathcal U^a\mathcal U^b h_{\mu\nu}$, the action becomes
\begin{equation}\label{eq:CW-hbar-cD}
 S_{\text{MNS}}[h;\cU]=\frac{\kappa}{2}\int_\Sigma d^2\sigma\,
 \mathcal U^a\mathcal U^b
 \left[
 h_{ss}D_asD_bs
 +2h_{si}D_as\,\partial_bY^i
 +h_{ij}\partial_aY^i\partial_bY^j
 \right].
\end{equation}
The above action is written only in terms of $\chi$-invariant quantities and is ready for our next step. 

\subsubsection{Eliminating \texorpdfstring{$s$}{s} and CW gauge fields}\label{sec:s-cW-elimination}

While \eqref{eq:CW-hbar-cD} is explicitly written in terms of $\chi$-invariant quantities, as \eqref{s-Wa-Y-chi} shows, $s, \cW_a$ which are $\chi$-noninvariant quantities, implicitly appear in the action. $\delta_\chi s=\chi$ implies that $s$ can be local gauged away. {Only $\cU^a\cW_a$} appears in the action; this combination is then eliminated by its algebraic equation, not by a second independent CW gauge choice. 

To eliminate $s, \cU^a \cW_a$, we note that if $h_{ss}\neq 0$, \eqref{eq:CW-hbar-cD} is algebraic in $\cW_a$ and quadratic in $\cU^a \cW_a$ and therefore,
\begin{equation}\label{eq:CW-hbar-cD-2}
 S_{\text{MNS}}[h;\cU]=\frac{\kappa}{2}\int_\Sigma d^2\sigma\,
 \left[ h_{ss} (\cU^a \cW_a+P_s)^2+ \mathcal U^a\mathcal U^b G_{ij}\partial_aY^i\partial_bY^j
 \right]
\end{equation}
where
\begin{equation}
    P_s:=\cU^a\partial_a s+ \frac{h_{si}}{h_{ss}} \cU^a\partial_a Y^i, \qquad  G_{ij}:=h_{ij}-\frac{h_{si}h_{sj}}{h_{ss}}.
\end{equation}
Upon equation of motion for $\cW_a$, $\cU^a \cW_a=-P_s$ and hence one remains only with, 
\begin{equation}\label{eq:CW-hbar-cD-W-replaced}
 S_{\text{MNS}}[h; \cU]
 =\frac{\kappa}{2}\int_\Sigma d^2\sigma\,
 \mathcal U^a\mathcal U^b G_{ij}(Y)
 \partial_aY^i\partial_bY^j,
\end{equation}
We note that $s$ has been completely and consistently eliminated from the {classical} action. Eliminating $\cU^a\cW_a$ {gives} the {same} {quadratic term, but quantum equivalence also requires accounting for the functional measure, determinants and gauge fixing}. We also note that $G_{ij}$ is a nondegenerate metric on a $(D-1)$-dimensional manifold $\bM$, which is the projective boundary of $\cM$, as {$G_{ij}$} is the metric on {the} local {space of dilation orbits}, the projective boundary $\bM$. {Identification with an asymptotic boundary requires a separate endpoint or conformal-completion map, as in the AdS example below.} Importantly, we note that $\delta_\chi G_{ij}=0$ and hence the right-hand-side (RHS) of \eqref{eq:CW-hbar-cD-W-replaced} is written in terms of quantities that are $\chi$-invariant, namely $\cU^a, Y^i, G_{ij}$.

We can now rewrite \eqref{eq:CW-hbar-cD-W-replaced} in a more explicit and inspiring way, as the main result of this section. For $\rho^2\neq 0$,
\begin{equation}\label{S-g-S-ILST-G}
 \inbox{
 S_{\rm MNS}[g;\mathcal V]
 =S_{\rm ILST}[G;\mathcal U] 
 }
\end{equation}
Eq.\eqref{S-g-S-ILST-G} shows upon eliminating $\cW_a, s$ using the $\cW_a$ EoM, the null string action on $D$ dimensional background $g_{\mu\nu}$ takes the form of ILST action on $D-1$ dimensional background with metric $G_{ij}$. Both sides of the equality \eqref{S-g-S-ILST-G} are strictly and explicitly invariant under all three gauge symmetries. Nonetheless, the RHS is written in terms of $\chi$-invariant quantities and hence one need not worry about dealing with the $\chi$ symmetry, {no additional CW constraint remains to be imposed}. Explicitly, the RHS may be treated as a standard ILST theory. To summarize this part, 
we have established that The ILST theory on background $\bM$ with metric $G_{ij}$ is classically equivalent to null string theory on background $\cM$ with metric $g_{\mu\nu}$. We conjecturally, promote this classical equivalence to full quantum level, that we state as,
\begin{center}
\textit{\textbf{Null String Holography conjecture:}}

\hspace*{-3mm}\inbox{{\textit{
 \hspace*{-4mm}Null string theory on background $\cM$ with metric $g_{\mu\nu}$  $\equiv$ 
 ILST theory on background $\bM$ with metric $G_{ij}$.\hspace*{-4mm}}}}
\end{center}

We crucially note that to arrive at the above statement we assumed $\cM$ admits a homothetic vector field $\rho^\mu\partial_\mu$, which in appropriate coordinate system is along the ``holographic direction'' $s$;  $\rho^\mu\partial_\mu=\partial_s$. Moreover, we assumed that the above is written for patches of $\cM$ in which the causal character of $\partial_s$ does not change; it remains spacelike or timelike over the patch we consider. The ILST side is then written on the projective boundary of the same patch of $\cM$. In general $\partial_s$ may change causal character over global $\cM$, then we have a patchwise null string holography for each patch with a given norm of $|\partial_s|$. {The statement} above {applies to} timelike or spacelike $\partial_s$. {For an everywhere-null flow, the connection instead remains a constraint multiplier, as we show next}.

\subsection{Null dilation orbits: \texorpdfstring{$\rho_\mu\rho^\mu=0$}{rho squared=0}}\label{sec:rho=0}

The preceding analysis relied critically on the assumption $\rho^2\neq 0$, which allowed the algebraic elimination of the ``holographic direction'' $s$ and the CW gauge field through its equation of motion. This procedure becomes singular when  the holographic direction $\partial_s$ is null, $\rho^2=0$. To see this, let us return to the action \eqref{eq:CW-g-cD},
\begin{equation}
S_{\text{MNS}}[g; \cV]=\frac{\kappa}{2}\int d^2\sigma\,
\left[
g_{\mu\nu}v^\mu v^\nu
+2w\,\rho_\mu v^\mu
+w^2\rho^2
\right].
\end{equation}
where,
\begin{equation}
v^\mu:=\mathcal V^a\partial_aX^\mu,
\qquad
w:=\mathcal V^a\mathcal W_a.   
\end{equation}
If $\rho^2\neq 0$, the $w$ terms can be made a complete square and hence $w$ may be removed upon its EoM (classically) or integrated out (quantum mechanically). This yields $w=- \frac{1}{\rho^2} \rho_\mu v^\mu$ and thus, $w$ is eliminated in the on-shell action. If $\rho^2=0$, however, the action is linear in $w$. EoM for $w$ in this case is, 
\begin{equation}\label{eq:null-constraint}
{\rho_\mu\mathcal V^a\partial_aX^\mu=0,} 
\end{equation}
which is a constraint among derivatives of $X^\mu$, rather than determining $w$. 

We now focus on a region where the dilation vector is everywhere null and nonvanishing, $\rho^2=0$ and $\rho^\mu\neq0$. We take the target metric to be Lorentzian and all fields to be real. As in the non-null case, we work locally where the dilation orbits form a smooth quotient and the worldsheet kernel density is nonvanishing. The aim is again to remove the coordinate along the dilation orbits and express the theory in terms of the remaining CW-invariant variables. The difference is that the connection equation now constrains the projected motion, rather than determining the connection. To this end, we again adopt the $(s,Y^i)$, with $s$ adapted to the dilation flow $\rho=\partial_s$ and $i=1,\ldots,D-1$. Eq.\eqref{Dilatation-Eq} yields, $g_{\mu\nu}=e^{2s}h_{\mu\nu}$ with $\partial_s h_{\mu\nu}=0$, while nullness implies $h_{ss}=0$. Therefore, 
\begin{equation}
h_{\mu\nu}:=e^{-2s}g_{\mu\nu},
\qquad
h_{ss}=0, \qquad h_{si}=h_{is}=h_{is}(Y),\quad h_{ij}=h_{ij}(Y)
\end{equation}

The one-form $h_{is}\dd Y^i$ is nowhere vanishing by
nondegeneracy of the ambient metric; no exactness or
hypersurface-orthogonality assumption is required.

Using the same CW-invariant variables as before,
\begin{equation}
 \mathcal U^a=e^s\mathcal V^a,
 \qquad
 B_a=\partial_as+\mathcal W_a,
\end{equation}
the action becomes
\begin{equation}
 S_{\rm MNS}[h;\mathcal U]
 =
 \frac{\kappa}{2}\int_\Sigma\dd^2\sigma\,
 \mathcal U^a\mathcal U^b
 \left[
 h_{ij}\partial_aY^i\partial_bY^j
 +2B_a h_{is}\partial_bY^i
 \right].
 \label{eq:null-orbit-action}
\end{equation}
The coordinate $s$ appears only through $B_a$ and can be
removed locally by the CW gauge choice $s=0$. The resulting
theory is defined on the $(D-1)$-dimensional orbit space
$\mathcal B_\rho$, with the connection retained as a
Lagrange multiplier.

Although $h_{ij}$ need not be degenerate, the kinetic term
can be expressed using a Carrollian metric representative.
Choose a vector field $V^i(Y)$ normalized by $h_{is}V^i=1$
and define
\begin{equation}
\begin{aligned}
 A_i
 &:=h_{ij}V^j
 -\frac12\bigl(h_{kl}V^kV^l\bigr)h_{is},
 \\
 \gamma_{ij}
 &:=h_{ij}-h_{is}A_j-h_{js}A_i.
\end{aligned}
\label{eq:null-Carroll-decomposition}
\end{equation}
Then $\gamma_{ij}V^j=0$, while $\gamma_{ij}$ agrees with
$h_{ij}$ on the subspace spanned by vectors $n^i$, {such that} $h_{is}n^i=0$, where the metric
is positive definite. Thus $\gamma_{ij}$ is positive
semidefinite of rank $D-2$: it measures spatial
displacements, with $V^i$ defining its degenerate
Carrollian time direction. This gives a Carrollian metric
structure on $\mathcal B_\rho$ for each normalized choice
of $V^i$.\footnote{
The vector $V^i$ is an auxiliary choice on the quotient,
not the projection of $\rho$, which is removed by
quotienting. This construction is distinct from the
Carroll geometry induced on ambient null hypersurfaces,
whose null generators are $\rho$ when
$h_{is}\dd Y^i$ is hypersurface orthogonal.
}

The invertible connection shift
\begin{equation}
 \widehat B_a:=B_a+A_i\partial_aY^i
 \label{eq:null-Carroll-connection}
\end{equation}
rewrites \eqref{eq:null-orbit-action} exactly as
\begin{equation}
 S_{\rm MNS}[h;\mathcal U]
 =
 \frac{\kappa}{2}\int_\Sigma\dd^2\sigma\,
 \mathcal U^a\mathcal U^b
 \left[
 \gamma_{ij}\partial_aY^i\partial_bY^j
 +2\widehat B_a h_{is}\partial_bY^i
 \right].
 \label{eq:null-Carroll-action}
\end{equation}
All quantities in this expression are CW invariant.
Changing the normalized vector $V^i$ changes
$\gamma_{ij}$ and $\widehat B_a$ in compensating ways,
leaving the action unchanged. The Carrollian metric
representative therefore depends on the decomposition,
while the full constrained theory does not.

The multiplier term preserves the CW constraint.
Writing $q^i:=\mathcal U^a\partial_aY^i$, variation of
$\widehat B_a$ and contraction of the $\mathcal U^a$
equation with $\mathcal U^a$ give
\begin{equation}
 h_{is}q^i=0,
 \qquad
 \gamma_{ij}q^iq^j=0.
 \label{eq:null-Carroll-constraints}
\end{equation}
Positivity of $\gamma_{ij}$ implies $q^i=f V^i$, and
the first constraint then sets $f=0$. Each worldsheet
generator therefore projects to a fixed point of
$\mathcal B_\rho$, although spatial string profiles
and the remaining multiplier equations survive.
The degenerate kinetic term alone would not impose
this restriction.
\subsection{Nearest to minimal null string backgrounds}\label{sec:NMNS-K1-K2}

So far we discussed the minimal null string formulation assuming that the background manifold $\cM$ admits a homothety vector. Even the simplest maximally symmetric manifolds like AdS or dS do not fulfill this requirement. So, we need to extend the previous construction to admit more physically interesting cases. To this end, we show that one can still define null string theory, essentially in the same manner, on manifolds that admit any CKV, not necessarily a homothety. 

Let manifold $(\mathcal M,[\cG ])$  admit at least one CKV, $K_\mu$, that are solutions to 
\begin{equation}\label{CK-Eq}
 \mathcal L_K\cG_{\mu\nu}=2\omega_K(X)\cG_{\mu\nu},\qquad \omega_K(X)={\frac{1}{D}}\nabla\cdot K, 
\end{equation}
where $\mathcal L_K$ denotes Lie-derivative along vector $K^\mu$ and  $\omega_K(X)$ is a  scalar function. 
Solutions to the conformal Killing equation \eqref{CK-Eq}, generically, fall into three categories:
\begin{enumerate}[leftmargin=4mm, itemsep=-1mm] 
\vspace*{-3mm}    \item Isometries, with $\omega_K(X)=0$
    \item Homothety, $\rho^\mu$, with $\omega_K(X)=1$ or $\nabla_\mu \rho^\mu={D}$
    \item Generic CVK with $\omega_K$ is a function of $X$.
\end{enumerate}
\vspace*{-3mm}
For {$D>2$,} {in} flat Minkowski space} there are $(D+1)(D+2)/2$ CKVs. AdS or dS, while have isometries, do not admit a homothety. 

To gauge the flow generated by a generic CKV $K$, we introduce a Carroll--Weyl connection
$\mathcal W_a$ and define
\begin{equation}
 D_aX^\mu
 :=\partial_aX^\mu+\mathcal W_aK^\mu(X).
 \label{eq:homothetic-covariant-derivative}
\end{equation}
The ``nearest to minimal null string'' (NMNS) action is
\begin{equation}
 S_{\rm NMNS}[\cG; \cE; K ]
 =\frac{\kappa}{2}\int_\Sigma \mathrm d^2\sigma\,
 \mathcal E^a\mathcal E^b \cG_{\mu\nu}(X)
 D_aX^\mu D_bX^\nu.
 \label{eq:CW-action-G-frame}
\end{equation}
It is invariant under
\begin{equation}
 \delta_\chi X^\mu=\chi K^\mu(X),
 \qquad
 \delta_\chi\mathcal W_a=-\partial_a\chi,
 \qquad
 \delta_\chi\mathcal E^a=-\chi\omega_K(X)\mathcal E^a.
 \label{eq:CW-transformations-G-frame}
\end{equation}
The above clarifies the name NMNS: the $\delta_\chi\mathcal V^a$ in the above has $X$ dependence for $\omega_K(X)\neq 1$. Explicitly, by the definitions and arguments in \cite{Sheikh-Jabbari:2026kwr}, for the \textit{minimal} null string action the CW transformations should be only a worldsheet feature and not theory and $X$-dependent. Moreover, {for isometries $\omega_K(X)=0$, so} the symmetry is a local $K$-shift with invariant $\mathcal E^a$ in this frame. On patches with $K\neq0$, the CW transformations above can be put in the normalized form, using the conformal and field redefinitions below. 

To proceed, we note that any CKV $K^\mu$ of metric $\cG_{\mu\nu}$ can be mapped to a homothety vector of a new metric $g_{\mu\nu}$ upon a simple scaling. To see this, consider
\begin{equation}
 g_{\mu\nu}:=e^{-2\Omega(X)}\cG_{\mu\nu}.
 \label{eq:target-conformal-frame}
\end{equation}
Its Lie derivative along $K$ is
\begin{equation}
 \mathcal L_Kg_{\mu\nu}
 =2\bigl(\omega_K-\Omega_K\bigr)g_{\mu\nu},
 \qquad
 \Omega_K:=K^\mu\partial_\mu\Omega.
 \label{eq:conformal-frame-Lie-derivative}
\end{equation}
On a patch where $K\neq0$, the first-order equation $\Omega_K=\omega_K-1$ can always be integrated locally along the orbits of $K$ to obtain $\Omega$. The existence of a smooth global solution 
to this equation is a separate question; here we work on a regular patch where such a frame exists. In this conformal frame,
\begin{equation}
 \mathcal L_Kg_{\mu\nu}=2g_{\mu\nu},
 \label{eq:target-homothety}
\end{equation}
so $K$ is a homothetic vector  of $g_{\mu\nu}$. The normalization in \eqref{eq:target-homothety} fixes the parametrization of its flow.  

One can now follow the steps of the previous section with $g_{\mu\nu}$. Define, 
\begin{equation}
 \mathcal V_K^a:=e^{\Omega(X)}\mathcal E^a.
 \label{eq:U-field-redefinition}
\end{equation}
Using \eqref{eq:CW-transformations-G-frame}, one obtains
\begin{equation}
 \delta_\chi\mathcal V_K^a
 =e^{\Omega}\left(
 \delta_\chi\mathcal E^a
 +\chi \Omega_K\mathcal E^a\right)
 =-\chi\mathcal V_K^a,
 \label{eq:intrinsic-U-transformation}
\end{equation}
and
\begin{equation}
 \cE^a \cE^b\ \cG_{\mu\nu}= \mathcal V_K^a\mathcal V_K^b\ g_{\mu\nu}.
 \label{eq:conformal-frame-cancellation}
\end{equation}
Consequently, 
\begin{equation}\label{GKE--gVK}
S_{\rm NMNS}[\cG; \mathcal E;K] =S_{\rm MNS}[g;\mathcal V_K],
\end{equation}

We note that the $K$-dependence in the RHS besides the $\cV_K$, which is written explicitly, is also carried in $g$, as the conformal factor $e^\Omega$ relating it to $\cG$ depends on $K$, cf. \eqref{eq:conformal-frame-Lie-derivative}. Given \eqref{GKE--gVK} we can run again the machinery worked out in section \ref{sec:CW-reduction}.
If $\cM$ has more than one CKV, the question  arises is which of these Killing vectors one should gauge to obtain a null string theory. The above construction would work for every choice of $K$; depending on the $K$ one takes, one may get a different null string theory, which one may call $K$- minimal null string theory, $K$-MNS. $K$-MNS is not a minimal theory, as it involves another vector field on $\cM$, it is a theory in ``nearest to minimal null string'' family. There are other ways to move away from minimality, that we briefly discuss in the discussion section. 

We close this section by three comments: (1) {on regular $\rho\neq0$ patches,} the null string holographic conjecture {extends} to the nearest minimal null string theory. The $K$-MNS theory on $\cM$ is {classically} equivalent to ILST on the space modded out by the orbits of  CKV $K^\mu$. {For an everywhere-null $K$, the clock-dependent multiplier action of section~\ref{sec:rho=0} must instead be retained.} The argument just follows from \eqref{GKE--gVK}. (2) This equivalence is patchwise, the metric $g$  typically covers a patch of the original space $\cG$ and one should give a patchwise description of the null strings in the ``holographic'' ILST side. (3) The null string holography statement above, was argued for in the cases where the target space is non-degenerate, and in particular non-Carrollian. Nonetheless, one can show a similar holographic statement also works for Carrollian target spaces \cite{Sheikh-Jabbari:2026nez}.

\section{Examples of null string holography}\label{sec:examples}

Having worked out the general formulation for (nearest to) minimal null strings on $\cM$ and its equivalent ILST theory on $\bM$, in this section we work out three examples with $\cM$ being AdS$_D$, $D$ dimensional Minkowski  and dS$_D$ spacetimes and illustrate how the null string holography works.

\subsection{Null strings on AdS in Poincar\'e patch}\label{sec:ads-reduction}

AdS space provides a classic example of the construction in
section~\ref{sec:CW-reduction}. Our principle/basic construction is based on the homothetic vector field  which in our adopted holographic coordinate is $\partial_s$. It is, however, known that AdS space does not admit a homothety, while admits CKVs. We should hence utilize the nearest minimal construction in section \ref{sec:NMNS-K1-K2} in which we need to identify which CKV $K$ we gauge. Different patches on AdS space can be covered by different slicings/coordinate systems and each make a CKV more manifest. Moreover, as discussed, there could be subtleties regarding whether $K^2=K^\mu K_\mu$ changes causal character over a given patch. In this section, for illustrative purposes, we only focus on the AdS in Poincar\'e patch in which the associated CKV is spacelike.

Metric on the Poincar\'e patch of $\mathrm{AdS}_{D}$ \cite{Aharony:1999ti, Bayona:2005nq}, written in terms of the physical metric $\mathcal G$ (cf. discussions and conventions in section \ref{sec:NMNS-K1-K2}), is
\begin{equation}
 \mathrm ds^2_{\mathcal G}
 =\frac{L^2}{z^2}
 \left(\mathrm dz^2+\eta_{ij}\mathrm dx^i\mathrm dx^j\right),
 \label{eq:AdS-Poincare-metric}
\end{equation}
where $i=0,\ldots,D-2$, $L$ is the AdS radius, and the coordinates $z$ and $x^i$ are dimensionless. The conformal boundary is at $z=0$, and the
Poincar\'e coordinates select the Minkowski representative $\eta_{ij}$ of
its conformal metric. On the patch $z>0$, the vector field
\begin{equation}
 \rho^\mu\partial_\mu=\partial_z
 \label{eq:AdS-radial-CKV}
\end{equation}
is a conformal Killing vector of the physical metric, satisfying
\begin{equation}
\mathcal L_\rho \mathcal G=-\frac{2}{z}\mathcal G,
\qquad
\mathcal G(\rho,\rho)=\frac{L^2}{z^2}>0,
\end{equation}
but it is not a homothety. To obtain a homothetic representative $g$
\begin{equation}
 g_{\mu\nu}=z^2e^{2z}\mathcal G_{\mu\nu}
 \label{eq:AdS-homothetic-g}
\end{equation}
Then
\begin{equation}
\mathcal L_\rho g=2g,
\qquad
g(\rho,\rho)=L^2e^{2z}>0,
\end{equation}
so $\rho$ is the required homothety and $z$ is the holographic direction ($s$ in the notation of section \ref{sec:ILST-Bulk}).

The integral curves of $\rho$ keep $x^i$ fixed and vary only $z$. Every such
curve has a distinguished endpoint at $z=0$, labelled by the same $x^i$.
Consequently, the local orbit space of this foliation is naturally
identified with the Minkowski patch of the AdS conformal boundary,
\begin{equation}
 \mathcal B_\rho:=\mathrm{AdS}_{D}^{\mathrm{Poincar\acute e}}/\mathrm{Flow}(\rho)\simeq\mathbb M_{D-1}.
 \label{eq:AdS-orbit-boundary}
\end{equation}
This endpoint map is the additional geometric input that turns the
homothetic quotient into a genuine boundary description.
The boundary identification refers to the conformal completion of the physical metric $\mathcal G$.

With the homothetic representative $g$ \eqref{eq:AdS-homothetic-g}, the null string action is
\begin{equation}
 S_{\mathrm{MNS}}[g; \cV]=\frac{\kappa L^2}{2} \int_\Sigma\mathrm d^2\sigma\, e^{2z}\mathcal V^a\mathcal V^b \left[(\partial_a z+\mathcal W_a)(\partial_b z +\mathcal W_b) +\eta_{ij}\partial_ax^i\partial_bx^j \right].
 \label{eq:AdS-CW-action-homothetic-frame}
\end{equation}
Its gauge transformations are
\begin{equation}
 \delta_\chi z=\chi,\qquad
 \delta_\chi x^i=0,\qquad
 \delta_\chi\mathcal W_a=-\partial_a\chi,\qquad
 \delta_\chi\mathcal V^a=-\chi\mathcal V^a.
 \label{eq:AdS-U-transformation}
\end{equation}
In the $h$-frame, where $h=e^{-2z}g$ and $\cU^a= e^z \cV^a$, 
\begin{equation}
 S_{\mathrm{MNS}}[h; \cU]= \frac{\kappa L^2}{2}
 \int_\Sigma\mathrm d^2\sigma\,
\left[
 (\cU^a\partial_a z+w)^2
 +\mathcal U^a\mathcal U^b\eta_{ij}\partial_ax^i\partial_bx^j
 \right],\qquad w:= \cU^a\cW_a.
 \label{eq:AdS-action-cU}
\end{equation}
On a worldsheet patch with $\mathcal U^a\neq0$, the $\mathcal W_a$ equation of motion
$\mathcal U^a(\mathcal U^b\partial_b z+w)=0$ gives $w=-\mathcal U^a\partial_a z$, and hence
\begin{equation}
\left.S_{\mathrm{MNS}}[h; \cU]\right|_{w=-\cU^a\partial_a z}
 =
 \frac{\kappa L^2}{2}
 \int_\Sigma\mathrm d^2\sigma\,
 \mathcal U^a\mathcal U^b
 \eta_{ij}\partial_ax^i\partial_bx^j
 = S_{\rm ILST}[L^2\eta; \cU].
 \label{eq:AdS-reduced-action-U}
\end{equation}
Equivalently, defining $\widetilde{\mathcal U}^a = L\,\mathcal U^a$, we may write the reduced action as $S_{\rm ILST}[\eta; \widetilde{\mathcal U}]$. That is,
\begin{equation}
 \inbox{
 {\text{minimal null string on }}
 \mathrm{AdS}_{D}^{\mathrm{Poincar\acute e}}
 \quad\simeq\quad
 \text{ILST string on }\mathbb M_{D-1} .}
 \label{eq:AdS-CW-ILST-correspondence}
\end{equation}
This is a patchwise classical equivalence of reduced phase spaces for the selected radial flow and homothetic representative. It
implements explicitly the boundary interpretation of the quotient discussed
in section~\ref{sec:CW-reduction}.
\color{black}

\subsection{Example II: de Sitter space}
\label{subsec:brief-dS-comparison}

In the AdS example we considered AdS in Poincar\'e patch where the gauged CKV is spacelike  spacelike over the patch. As the next example, we consider dS space in cosmological patch which provides an example for timelike CKV  field.   Metric on $\mathrm{dS}_{D}$ in conformal cosmological patch is,
\begin{equation}
 \mathrm ds_\cG^2
 =\frac{1}{H^2\eta^2}\left(-\mathrm d\eta^2+\delta_{ij}\mathrm dx^i\mathrm dx^j\right).
 \label{eq:dS-conformal-patch}
\end{equation}
The boundary-adapted vector $\rho=\partial_\eta$, for which $\mathcal L_\rho \cG_{\mu\nu}=-\frac{2}{\eta}\cG_{\mu\nu}$ and $\rho^2<0$. Orbits of $\partial_\eta$ keep $x^i$ fixed hence the corresponding ILST theory should reside at spacelike constant $\eta=0$ slice. Since the analysis is basically the same as as the one in previous section we do not repeat here. The statement of null string holography for this case is, 
\bc
\textit{minimal null string theory on dS$_D$ in cosmological patch is described by the ILST action on Euclidean constant time slices of \eqref{eq:dS-conformal-patch}.}
\ec

For real fields, the positive-definite constant $\eta$surface metric implies $\mathcal U^a\partial_a x^i=0$, similarly to the timelike Minkowski sector of the next section. The transverse canonical momenta vanish on the constraint surface, where the quadratic null constraint is irregular; the usual regular-constraint counting of degrees of freedom does not apply. The limits of this conformal patch do not include the big-bang curvature singularity of de Sitter space. A global or cosmological patch description requires a separate {choice of flow and boundary analysis, which we postpone to future work}.

\subsection{Example III: Null strings on Minkowski spacetime}\label{sec:Mink-bdry}

As the third example, let us consider the Minkowski target space with metric $\eta_{\mu\nu}$, with signature $(-+\cdots+)$. It admits a homothety vector
\begin{equation}
\rho=X^\mu\partial_\mu,
\qquad
\mathcal L_\rho\eta_{\mu\nu}=2\eta_{\mu\nu}.
\end{equation}
As we see $\rho^2=X^\mu X_\mu$ does not have a definite sign. One can distinguish three cases, $X^2\gtreqless 0$.

The asymptotic boundary can be described by five codimension-one pieces,
\begin{equation}
\mathcal B_{\mathrm{asym}}
={\mathcal S}\cup{\mathcal T^+}
\cup{\mathcal T^-}\cup{\mathcal I^+}
\cup{\mathcal I^-},
\end{equation}
where
\begin{equation}
\begin{aligned}
\mathcal S
&\simeq\mathrm{dS}_{D-1},\\
\mathcal T^\pm
&\simeq\mathbb H^{D-1},\\
\mathcal I^\pm
&\simeq\mathbb R\times S^{D-2}.
\end{aligned}
\end{equation}
Here $\mathcal S$ and $\mathcal T^\pm$ describe the spacelike and timelike approach directions to $\iota^0$ and $\iota^\pm$, respectively, while $\mathcal I^\pm$ are future and past null infinity \cite{Compere:2023qoa}. Each piece has dimension $D-1$, with adjacent pieces joined
along their common limiting boundaries. 
The non-null pieces coincide with the corresponding dilation quotients. At null infinity, however, we retain the retarded or advanced time, so $\mathcal I^\pm$ remains a Carrollian hypersurface. The three cases $X^2<0$ (associated with $\mathcal T^\pm$), $X^2>0$ (associated with ${\cal S}$) and $X^2=0$ (associated with ${\cal I}^\pm$) should be analyzed separately.
\color{black}

\paragraph{Spacelike dilation orbits: $X^2>0$.}
Consider first the region $X^2>0$, in which the Euler vector is spacelike. Parameterize
\begin{equation}
X^\mu=e^sY^\mu,
\qquad
Y^2=1,
\end{equation}
where  by definition $Y^\mu$ are embedding coordinates for a unit radius dS$_{D-1}$. It may be solved through $y^A, A=1,\ldots,D-1$. The flat metric becomes
\begin{equation}
\eta_{\mu\nu}\mathrm dX^\mu\mathrm dX^\nu
=
e^{2s}\left[
\mathrm ds^2+
g^{\mathrm{dS}}_{AB}(y)\mathrm dy^A\mathrm dy^B
\right].
\end{equation}
The invariant action, on solutions for the CW gauge connection, is immediately obtained as
\begin{equation}
S_{\mathrm{MNS}}[h;\mathcal U]
=\frac{\kappa}{2}\int\mathrm d^2\sigma
\mathcal U^a\mathcal U^b
g^{\mathrm{dS}}_{AB}\partial_ay^A\partial_by^B
\end{equation}
Thus, on $\mathcal U^a\neq0$,
\begin{equation}
\inbox{
S_{MNS}[\eta; \cV; X^2>0]=S_{\mathrm{ILST}}[g^{\mathrm{dS}};\mathcal U]}
\end{equation}
The quotient metric is Lorentzian, so nonzero null propagation remains possible. This is the flat-ambient/de Sitter reduction identified in \cite{Lindstrom:2026tua}.

\paragraph{Timelike dilation orbits: $X^2<0$.}
For $X^2<0$, parameterize
\begin{equation}
X^\mu=e^sY^\mu,
\qquad
Y^2=-1,
\qquad
\pm Y^0>0.
\end{equation}
There are two disconnected unit hyperboloids $\mathbb{H}^{D-1}$, distinguished by the sign of $Y^0$. Since positive scaling does not reverse the time orientation, the future and past sectors remain separate. In these coordinates,
\begin{equation}
\eta_{\mu\nu}\mathrm dX^\mu\mathrm dX^\nu
=
e^{2s}\left[
-\mathrm ds^2+
g^{\mathbb{H}}_{AB}(y)\mathrm dy^A\mathrm dy^B
\right],
\end{equation}
and, on solutions to CW gauge connection, 
\begin{equation}
S_{\mathrm{MNS}}[h;\mathcal V; X^2<0]
=\frac{\kappa}{2}\int\mathrm d^2\sigma
\mathcal U^a\mathcal U^b
g^{\mathbb{H}}_{AB}\partial_ay^A\partial_by^B
\end{equation}
yielding 
\begin{equation}
\inbox{
S_{MNS}[\eta;\cV; X^2<0]=S_{\mathrm{ILST}}[g^{\mathbb{H}};\mathcal U],
}
\end{equation}
Since $g^{\mathbb{H}}$ is positive definite,
\begin{equation}
g^{\mathbb{H}}_{AB}v^Av^B=0, \qquad v^A:=\mathcal U^a\partial_ay^A
\quad\Longrightarrow\quad v^A=0
\end{equation}
for real fields. Spatial profiles can remain, but they do not evolve along the worldsheet direction selected by $\mathcal U^a$. This is an ILST action with the Euclidean target $\mathbb{H}^{D-1}$, without transverse null propagation.

\paragraph{Null dilation orbits: $X^2=0$, $X\neq0$.}

The dilation vector $\rho=X^\mu\partial_\mu$ is null only on the light cone, $X^2=0$. This differs from the setting of section~\ref{sec:rho=0}, where the chosen vector is null throughout a region of the target spacetime. The non-null reduction divides by $\rho^2=X^2$, so its formulas cannot be used by simply setting $X^2=0$.

One possibility is to take the light cone itself as the string target. Writing $X^\mu=e^s n^\mu(y)$ with $n^2=0$ and $n^0=\pm1$, the induced metric and dilation vector are
\begin{equation}
h_{\rm cone}=e^{2s}\gamma_{AB}(y)\dd y^A\dd y^B,\qquad
K=\partial_s,\qquad h_{\rm cone}(K,\cdot)=0,\qquad
\mathcal L_Kh_{\rm cone}=2h_{\rm cone},
\end{equation}
where $\gamma_{AB}$ is the unit $S^{D-2}$ metric and $K$ is $\rho$ restricted to the cone. This is the Carrollian target space considered in \cite{Sheikh-Jabbari:2026nez}: the CW gauge field drops out, and the metric-only action is already CW invariant. The redefinition $\mathcal U^a=e^s\mathcal V^a$ removes $s$ from the action, leaving ILST on $S^{D-2}$. Given the above, one may extend the null string holography to the cases with Carrollian targets: The minimal null string theory and ILST on such spaces are (classically) equivalent \cite{Sheikh-Jabbari:2026nez}.

For comparison, consider a different choice of $K$ that is null everywhere in flat spacetime. In null Cartesian coordinates, take
\begin{equation}
\dd s_{\mathcal G}^2=2\dd s\dd v+\delta_{AB}\dd y^A\dd y^B,
\qquad K=\partial_s,\qquad A=1,\ldots,D-2.
\end{equation}
Here $K$ generates translations along $s$ and is a Killing vector of $\mathcal G$. Following section~\ref{sec:NMNS-K1-K2}, we choose $\Omega=-s$ and $g=e^{2s}\mathcal G$. The remaining target coordinates are $Y^i=(v,y^A)$, with $h_{is}\dd Y^i=\dd v$. Choosing $V=\partial_v$ in \eqref{eq:null-Carroll-decomposition} gives $A_i=0$, $\widehat B_a=B_a$ and the flat transverse metric $\delta_{AB}$.
With
\begin{equation}
\mathcal U^a=e^s\mathcal V_K^a=\mathcal E^a,
\qquad B_a=\partial_as+\mathcal W_a,
\end{equation}
the reduced action is
\begin{equation}
S_{\rm red}^{K}=\frac{\kappa}{2}\int_\Sigma\dd^2\sigma\,
\mathcal U^a\mathcal U^b
\left[\delta_{AB}\partial_a y^A\partial_b y^B
+2B_a\partial_bv\right].
\label{eq:Mink-null-infinity-action}
\end{equation}
This is the action \eqref{eq:null-Carroll-action} on a $(D-1)$-dimensional Carrollian target with metric $\delta_{AB}\dd y^A\dd y^B$, kernel vector $V=\partial_v$ and clock one-form $\dd v$. The $B_a$ equation imposes $\mathcal U^a\partial_a v=0$, and the remaining equations imply $\mathcal U^a\partial_a y^A=0$ for real fields. Spatial profiles can remain, subject to the other field equations. This example gauges a null translation, rather than the dilation $\rho=X^\mu\partial_\mu$. Identifying the resulting target with a patch of $\mathcal I^\pm$ requires a separate argument and is not established here.

\section{Discussion}  
\label{sec:conclusions}
Null string theory is the $2d$ worldsheet theory which realizes diffeomorphisms, Weyl and Carroll-Weyl (CW) scaling as gauge symmetries \cite{Sheikh-Jabbari:2026kwr}. One may rewrite the theory directly in terms of quantities that CW scaling invariant. We showed all degrees of freedom (dof), except the ``holographic direction'' $s$ and the CW gauge connection $\cW_a$, can be put in gauge invariant combinations. {On non-Carrollian target space}, $s$ can be removed locally by CW gauge fixing and the coupled component of $\cW_a$ by its algebraic equation and after eliminating/removing $s$, one remains with the ILST theory on the ``projective boundary'' of the original null string target space. We dubbed the equivalence between minimal null string theory on a manifold $\cM$ and the ILST theory on its projective boundary $\bM$, the \textit{null string holography}.

The vector along holographic direction $\partial_s$ can be spacelike, null or timelike. 
AdS space in Poincar\'e patch provides a prime example of spacelike $s$; dS in cosmological patch is an example of timelike $s$ and {the null-translation example in} flat Minkowski space {illustrates} the null case. {The reduced non-Carrollian metric} is {Euclidean/Lorentzian for a timelike/spacelike flow.

The null string holography {constructed} {here} is patchwise: Depending on the coordinate patch one adopts on the target space $\cM$, a CKV vector of $\cM$ is made manifest and may be gauged. Projective boundary depends on orbits of $K$. One the other hand, as discussed in section \ref{sec:NMNS-K1-K2}, depending on which $K$ one uses to gauge the CW scaling, one gets a different null string theory and the ``dual'' ILST theory knows about that through the projective boundary (along orbits of $K$). 
In other words, {the} theory and its reduction depend on {the selected} $K$. {Non-null orbits give ordinary ILST, whereas everywhere-null orbits give} the action discussed in section~\ref{sec:rho=0}. Prime example of this is the flat space $\mathbb{M}_D$. As discussed, one can use dS$_{D-1}$ slicing of flat space and gauge the CKV normal to the dS slice (which is a spacelike vector) to obtain ``null string theory on $\mathbb{M}_D$ is dual to ILST on dS$_{D-1}$'', see also \cite{Lindstrom:2026tua}. For the null sector of the projective boundary of $\mathbb{M}_D$, ${\cal I}^\pm$ pieces, we deal with null string theory and/or ILST theory on a Carrollian target space \cite{Sheikh-Jabbari:2026nez}.

We propose \emph{null string holography} as a quantum extension of
the classical  correspondence established here and its extension discussed in \cite{Sheikh-Jabbari:2026nez}. Its
verification requires compatible quantizations of the parent and
reduced theories, including their functional measures, gauge fixing,
and possible anomalies. The first steps in quantization of the gauged action has been taken in \cite{Rasulian:2026jvg, Duary:2026rlo, Duary:2026lmk, Chen:2026cau}, but a lot more has to be done, both in the minimal null string side and the ILST side. 

Here we mainly focused on the minimal null strings, while discussing briefly the nearest to minimal family. There are many different nonminimal null string theory families that may be constructed through additional worldsheet or target-space fields and symmetries. Such theories should be developed in their own right. The null string holography can be particularly applied to study strings probing black hole  or cosmological horizons \cite{ Bagchi:2023cfp,Bagchi:2024rje, Bagchi:2026wcu}. This analyses may shed light on  horizon degrees of freedom.

\paragraph{Acknowledgments.}
We thank Alireza Akbari, Arjun Bagchi, Aritra Banerjee, Daniel Grumiller,  Ida Rasulian and Shing Tung Yau for discussions. We especially thank Ulf Lindstr\"om, Bo Sundberg for sharing their paper on with very similar results prior to the arXiv submission. MMShJ acknowledges Iranian National Science Foundation (INSF) research chair grant No.~4045163. HY is supported in part by Beijing Natural Science Foundation under Grant No.~IS23013.

\end{document}